\documentclass[a4paper,12pt]{article}
\usepackage[utf8]{inputenc}
\usepackage[english]{babel}
\usepackage{amsmath}
\usepackage{amsfonts}
\usepackage{amssymb}
\usepackage{float,natbib}
\usepackage{graphicx}
\usepackage{geometry}
\usepackage{caption}
\usepackage{subcaption}
\usepackage{url}
\usepackage{booktabs}
\usepackage{dcolumn}
\usepackage{hyperref}
\usepackage{lscape,color}
 
\title{\Large Social networks, happiness and health: from sentiment analysis to a multidimensional indicator \\ of subjective well-being}
\author{\normalsize Stefano Maria Iacus
		\thanks{Department of Economics, Management and Quantitative Methods, Universit\`a degli Studi di Milano}
            \and
       	\normalsize Giuseppe Porro
       	\thanks{Department of Law, Economics and Culture, Universit\`a degli Studi dell'Insubria}
            \and
		\normalsize Silvia Salini
		\footnotemark[1]
            \and
        \normalsize Elena Siletti
        \footnotemark[1]
}

\date{}
\begin{document}
\maketitle

\paragraph{Abstract:}

This paper applies a novel technique of opinion analysis over social media data with the aim of proposing a new indicator of perceived and subjective well-being. This new index, namely SWBI,  examines several dimension of individual and social life. The indicator has been compared to some other existing indexes of well-being and health conditions in Italy: the BES (Benessere Equo Sostenibile), the incidence rate of influenza and the abundance of PM10 in urban environments. SWBI is a daily measure available at province level. BES data, currently available only for 2013 and 2014, are annual and available at regional level. Flu data are weekly and distributed as regional data and PM10 are collected daily for different cities. 
Due to the fact that the time scale and space granularity of the different indexes varies, we apply a novel statistical technique to discover nowcasting features and the classical latent analysis to study the relationships among them. A preliminary analysis suggest that the environmental and health conditions anticipate several dimensions of the perception of well-being as measured by SWBI. Moreover, the set of indicators included in the BES represent a latent dimension of well-being which shares similarities with the latent dimension represented by SWBI.

\paragraph{Keywords:}
Well-being, sentiment analysis, happiness, health

\paragraph{J.E.L. Classification:}
I10; I31

\section{Introduction}
\label{intro}

This paper proposes a set of well-being indicators, derived from a new supervised technique of web opinion analysis designed to capture several aspects  of subjective well-being from on-line discussions and then tries to relate these information with health indicators. The several dimensions of subjective well-being extracted from web conversations are aggregated into a unique index called SWBI (Social Well-Being Index). Our main purpose is to investigate whether SWBI and its single components - which are characterized by a daily frequency and a relatively low cost of acquisition - may adequately represent the reaction of a community to changes in health and environmental conditions.

Indeed, as \cite{steptoe_etal_2015} remarked, subjective well-being is definitely associated with health conditions. Recent literature has linked depression and life stress with premature mortality, coronary heart disease, diabetes, disability, and other chronic disorders \citep{steptoe_2006}.

Emotional states are seen as a key determinant of the somatization of feelings of stress and anxiety related to life events \citep{sabatini_2014}. Medicine and psychology studies suggest that emotional reactions to life events can affect physiology in ways that are damaging for health \citep{rozansky_etal_1999, kuhn_etal_2009}. But also the relationship between job loss, depression, and poor health conditions seems to be well established by several studies \citep{clark_oswald_1994, kuhn_etal_2009}. \cite{lyubomirsky_etal_2005} found that subjective well-being is a protective factor for health. \cite{chida_steptoe_2008} suggested that positive life evaluation and happiness predict lower future mortality and morbidity. Because of this documented relationship between well-being, health and quality of life, health care systems should be concerned not only with illness and disability, but also with supporting methods to improve psychological states \citep{steptoe_etal_2015}.

Unfortunately,  indicators of individual and social well-being are usually obtained from surveys based on self-evaluation of the respondents on life quality. Several studies have pointed out that explicitly asking for an evaluation of well-being generates biased estimates of the variable of interest. Social networks and web forums offer, on the contrary, a valuable, large and continuously updated source of spontaneous self-evaluations of life satisfaction, under many different aspects. Therefore, a new stream of studies is trying to extract reliable information from these media.

The SWBI  has been calculated on the 2012-2015 period in Italy and correlated to the available information on public health and other well-being indicators. Being that some of the dimension composing the SWBI capture the features of the interpersonal relation system, the analysis can shed some light on the reciprocal influence among social interactions, health and life quality perception, which is a necessary condition to deepen the study of causal relations among these variables.

The technique of opinion analysis used in this paper is the  iSA (integrated Sentiment Analysis) algorithm \citep{ceron_etal_2013,ceron_etal_2015} that extracts the sentiment from texts posted on social networks and has already been used to capture instantaneous happyness from social media data \citep{curini_etal_2015}.

An environment that is resilient and in a state of vitality is a prerequisite for a healthy and therefore happy life. Clean water, fresh air and uncontaminated food are only possible in a ``healthy" environment in which the size of naturalness is able to integrate with human activities \citep{bes_2014}.

The SWBI is then compared to some other existing indexes of well-being and health conditions in Italy: the BES (Benessere Equo Sostenibile), the incidence rate of influenza and the abundance of PM10 in urban environments. While SWBI is a daily measure available at province level, the BES data, currently available only for 2013 and 2014, are annual and available at regional level. Flu data are weekly and distributed as regional data and PM10 are collected daily for different cities. 
Due to the fact that the time scale and space granularity of the different indexes varies, we apply a novel statistical technique to discover ``nowcasting'' features and some type of latent analysis to study the relationships among them. A preliminary analysis suggest that the environmental and health conditions anticipate several dimensions of the perception of well-being as measured by SWBI. Moreover, the set of indicators included in the BES represent a latent dimension of well-being which shares similarities with the latent dimension represented by SWBI.

The paper is organized as follows: in Section \ref{sec:1} traditional measures of well-being are reviewed. Section \ref{sec:2} presents the new measure for subjective well-being. Section \ref{sec:3} is dedicated to methodological aspects of opinion analysis and nowcasting. Finally \ref{sec:4} present the applications of the proposed methodology analyzing SWBI along with other health and  social well-being indicators.

\section{Traditional measures of well-being}
\label{sec:1}

For a long time Gross Domestic Product (GDP) has been considered a good indicator of well-being. The reasons why GDP had such a success are its capacity to connect goods and services with different nature thanks to monetary valuation \citep{stiglitz_etal_2009}, its linear methodology, objectivity and clearness and the usefulness in international comparisons. Although its link with many standard of living indicators, is strong, this link is not universal and differences in income explain only a low proportion of the differences in happiness among peoples \citep{frey_stutzer_2002}: in fact, GDP has been criticized for being a weak indicator of social welfare and therefore a misleading guide for public policies \citep{fleurbaey_2009}.

In 2009, the so-called Stiglitz Commission \citep{stiglitz_etal_2009} observed that GDP should not be completely dismissed and proposed to build a complementary statistical system, centred on social well-being and suitable for measuring sustainability, composed by a wide set of indicators, quantitatively measured and representing both objective and subjective assessment of well-being, including also people's perception of their quality of life. With its work, the Commission made a sort of ``paradigmatic'' choice that have had a strong influence in further well-being literature and, above all, practice.

Following this path, statisticians and social scientists, governments and international organizations \citep{fleurbaey_2009} have developed a huge number of well-being indicators, with different structures, considering a great variety of dimensions and for many purposes. Examples of those indices are: the Human Development Index (HDI), the Better Life Index (BLI), the Happy Planet Index (HPI), the Fair Sustainable Well-Being Index (Benessere Equo Sostenibile-BES), the Canadian Index of Well-being and the Gross National Happiness Index (GNH).

After the Stiglitz Commission work, economists, psychologists and philosophers have become increasingly interested in self-reported measures of well-being. Among all the surveys used to study subjective well-being, different types can be highlighted: surveys of general nature that are submitted worldwide (Gallup World Poll, World Database of Happiness, World Values Survey), surveys of general nature that have local impact (Gallup-Healthways Well-being Index, British Household Panel Survey, European Social Values Survey(ESS), Eurobarometer, Global Health \& Wellbeing Survey), surveys that consider only certain groups of people, as youth (National Child Development Survey, Survey of Well-being of Young Children (SWYC)), students or employers (Gallup's surveys of workers and customers corporate clients,  Social-Emotional Well-being (SEW) Survey, GA Releases Graduate Student Happiness \& Well-Being Report).

In the last years self-reports are extensively used to study well-being, forgetting that they are often a misleading source of information \citep{schwarz_1999}. Reports of well-being are influenced by manipulations of current mood and of the immediate context, including earlier questions on a survey that cause particular domains of life to be temporarily salient \citep{schwarz_strack_1999}. Satisfaction with life and with particular domains is also affected by comparisons with other people and with past experiences \citep{clark_2003}. To overcome biases of self-reported measure of well-being \cite{kahneman_etal_2004} proposed some new procedures: the Experience Sampling Method (ESM), the Daily Reconstruction Method (DRM), and the Event Recall Method (ERM). The same scholars admit the limitations of these approaches: ESM, e.g., is not a practical method for national well-being accounts, because it is impractical to implement in large samples, the rate of nonresponse may be unacceptable and infrequent activities are only rarely sampled.

To solve the problem that individuals may interpret and use the response categories differently, survey researchers try to anchor response categories to words that have a common and clear meaning across respondents, but there is no guarantee that respondents use the scales comparably. Questions remain about whether one should give a cardinal interpretation to the numeric values attached to individual responses about their life satisfaction or emotional states because of the potential for personal use of scales \citep{kahneman_krueger_2006}.

Despite all the efforts made in the literature, and partly presented above, it remains much uncertainty in the use of self-reported data. Indeed, for example, in analyzing Gallup data Angus Deaton himself raised concerns on the use of social well-being questionnaire items for cross-national comparisons \citep{deaton_2012} as the order in which questions were asked may strongly affect the answers: specifically, shifting questions about politics  just before the questions on life evaluation increase the negative perception of the economic crisis.


This casts a shadow on the ability of the surveys, regardless of the sophisticated structure that may have, to grasp the structural component of life satisfaction. Since the surveys are instantaneous, they seem mainly suitable to capture the emotional, short-run component.

Summarizing and simplifying the issues about surveys that are still open, the two main limitations of the indicators based on self-reported information seem to be related to: a) the influence that the single question (or even the order the questions are put) can have on the answers quality; b) the limited frequency of the surveys, that may fail in capturing the trend changes in subjective well-being over time and in distinguishing between the short-run ``emotional'' component of well-being and the structural component, usually called ``life evaluation" or ``life satisfaction". The two components are presumably connected to different features of individual and collective life.

Economists typically watch what people do, rather than listening to what people say, as it has been argued by \cite{ditella_macculloch_2008}. Nevertheless, the development of Internet and, in particular, of a number of widespread social networks offers nowadays a rich source of information on public opinion, which is available without submitting any questionnaire or carrying out any systematic survey, they simply allow to listen to people. Social networks host an open, enormous amount of records and digital interactions \citep{pentland_2014}, under the form of microblogging, that can be collected and analyzed for research purposes, making it possible to study social dynamics from an unseen perspective. Analyzing this kind of data allows to listen to what people say: in well-being researches this means to be able to measure happiness in real-time, mapping its fluctuation due to the occurrence of external facts \citep{curini_etal_2015}.

\section{A new measure for well-being}
\label{sec:2}
Sentiment analysis is the core aspect of a brand new method for measurement of happiness and well-being. This research field is largely dedicated to the systematic extraction of web users' emotional state from the texts they post autonomously on different internet platforms, such as blogs, forums, social networks (e.g.,Twitter or Facebook) \citep{kramer_2010, ceron_etal_2013}. The availability of these large data sets have driven up the growth of theories and methodologies for  sentiment or opinion analysis. Despite many limitations \citep{Couper}, if correctly performed,  sentiment analysis seems to be a useful framework to exploit when the constraints of standard survey methodology may be too strong \citep{iacus2014}. On one hand, in fact, there is no need for asking questions to the target population: all that the analyst has to do is to listen to the on-line conversations and classify the opinions expressed accordingly; on the other hand, the available information is updated in real time and hence the frequency of the well-being evaluation can be as high as desired, theoretically allowing for separating the volatile and emotional component from the permanent and structural one.

Here we propose to apply iSA (integrated Sentiment Analysis) method to derive a set of indicators of subjective well-being that capture different aspects and dimensions of individual and collective life. The indicators are summarized in a global index named Social Well Being Index (SWBI). The term ``social" emphasizes that:
\begin{itemize}
\item the indicator monitors the subjective well-being expressed by the society through the social networks;
\item SWBI is not the result of some aggregation of individual well-being measurements: as it will be clear in what follows, the index directly measures the aggregate composition of the sentiment throughout the society.
\end{itemize}
The  iSA technique \citep{ceron_etal_2015}, that will be explained in details in Section \ref{sec:3}, has  been previously used to build a social media happiness indicator known as iHappy \citep{curini_etal_2015}.

\subsection{Aspects captured by the SWBI}
The eight indicators we evaluate concern three different well-being areas: personal well-being, social well-being, well-being at work. To be comparable with a composite well-being index currently available through periodical surveys for the main European countries, we adopt the definitions introduced by the think-tank NEF (New Economic Foundation). Each well-being area is analyzed by a single component and each component is defined through the hypothetic question one might find in a questionnaire \citep{nef_2012}. Let us point out once more that, in our case, these questions are just ``hypothetic": no explicit question can be submitted to the target population in our research, the sentiment and any kind of opinion are extracted from the text through the supervised analysis of the language used in the posts. The data source are tweets written in Italian language and from Itally and data are accessed through Twitter's public API. A small part of these data (around 1- to 5\% each day) contain geo-reference information which is used to build the SWBI indicator at province level in Italy. We have stored and analyzed more than 143 millions of tweets, about 100 thousands per day, of which only 1.2 millions of tweet are geo-localized at province level (about 1\%).

Here is the definition of each single components of SWBI:
\begin{enumerate}
\item {\it Personal well-being}:
\begin{itemize}
	\item {\bf emotional well-being}: the overall balance between the frequency of experiencing positive and negative emotions, with higher scores showing that positive emotions are felt more often than negative ones ($\tt emo$);
	\item {\bf satisfying life}: having positive evaluation of one's life overall ($\tt sat$);
	\item {\bf vitality}: having energy, feeling well-rested and healthy, and being physically active ($\tt vit$);
	\item {\bf resilience and self-esteem}: a measure of individual psychological resources, optimism and ability to deal with life difficulties ($\tt res$);
	\item {\bf positive functioning}: feeling free to choose and having the opportunity to do it; being able to make use of personal abilities and feeling absorbed and gratified in activities ($\tt fun$);
\end{itemize}
\item {\it Social well-being}:
\begin{itemize}
	\item {\bf trust and belonging}: trusting other people, feeling to be treated fairly and respectfully and feeling sentiments of belonging ($\tt tru$);
	\item {\bf relationships}: extent and quality of interactions in close relationships with family, friends and others who provide support ($\tt rel$);
\end{itemize}
\item {\it Well-being at work}:
\begin{itemize}
	\item {\bf quality of job}: feeling job satisfaction, satisfaction with work-life balance, evaluating the emotional experiences of  work and work conditions ($\tt wor$).
\end{itemize}
\end{enumerate}
As it is not possible to ask questions in social media, the components of the SWBI are obtained through the reading of a sample of tweets (see next Section for details) and trying to classify each tweet according to the scale -1, 0, 1, where -1 is for negative , 0 is neutral and 1 is positive feeling. For example, a text like ``I am grateful to my friends and relatives who sustained me during my hard times", will be classified as $\tt rel = +1$.
While a text like ``you can't really trust anyone nowadays", will be classified as $\tt tru = -1$; or a text like ``ok, let's go to work again today" as $\tt wor = 0$. These are of course just examples of how one derives the indicators from qualitative text analysis.

\section{How to analyze social media data}
\label{sec:3}
In this section we briefly present the iSA algorithm and the lead-lag estimation technique which are used in  Section \ref{sec:4.1} to, respectively, transform texts into opinions and to discover time-dependence between the proposed index and other well-being indexes.
\subsection{iSA}
iSA is a human supervised statistical method, where part of the texts are read by humans and part is classified by the machine. The supervised part is essential in that this is the step where information can be retrieved from texts without relying on dictionaries of special semantic rules. Human just read a text and associate a topic or opinion (for example: $D$ = ``satisfied at work'') to it. Then, the computer learn the association between the whole set of words used in a text to express that particular opinion and extends the same rule to the rest of the texts to be analyzed.

More formally, let us denote by $\mathcal D=\{D_0, D_2, \ldots, D_M\}$ the set of possibile categories (i.e. sentiments or opinions). The target of interest is $\{P(D), D\in\mathcal D\}$, i.e. the distribution of opinions in a corpus of $N$ texts. Normally, $D_0$ refers to  the texts corresponding to Off-topic or not relevant texts with respect to the analysis (i.e. the \textit{noise} in this framework).
Let $S_i$, $i=1, \ldots, K$, be a unique vector of $L$ possible stems (i.e. single words, unigrams, bigrams, etc which remain after the stemming phase) which identifies one of the texts in a corpus. More than one text in the corpus can be represented by the same $S_i$ and is such that each element of it is either $1$ if that stem is contained in a text, or $0$ in case of absence. The data set is then formalized as the set $\{ (s_j, d_j), j=1, \dots, N\}$ where $s_j\in\bar{\mathcal  S}$ (the space of possible vectors $S_j$) and $d_j$ can either be ``NA'' (not available or missing) or one of the hand coded categories $D\in \mathcal D$.

The ``traditional'' approach  includes all machine learning methods and statistical models that: 
\begin{enumerate}
\item use the individual hand coding from the training set to construct a model $P(D|S)$ for $P(D)$ as a function of $S$,  e.g. multinomial regression, Random Forests (RF), Support Vector Machines (SVM) etc.; 
\item predict the outcome of $\hat d_j=D$ for the texts with $S=s_j$ belonging to the test set; 
\item  when all data have been imputed in this way, these estimated categories $\hat d_j$ are aggregated to obtain a final estimate of $\hat P(D)$.
\end{enumerate}
 In matrix form, we can write
\begin{equation}
\underset{M\times 1}{P(D)} =  \underset{M\times  K}{P(D|S)} \underset{ K\times 1}{P(S)}
\label{eq1}
\end{equation}
where $P(D)$ is a $M\times 1$ vector, $P(D|S)$ is a $M\times  K$ matrix of conditional probabilities and $P(S)$ is a $ K\times 1$ vector which represents the distribution of $S_i$ over the corpus of texts. Once $P(D|S)$ is estimated from the training set with, say, $\hat P(D|S)$, then for each document in the test set with stem vector $s_j$, the opinion $\hat d_j$ is estimated with the simple Bayes estimator as the maximizer of the conditional probability, i.e. $\hat d_j= \arg\max_{D\in\mathcal D} \hat P(D|S=s_j)$. As it is well known, the present approach does not work if $P(D_0)$ is very large compared to the rest of the $D_i$'s.
iSA \citep{ceron_etal_2015} follow the idea by \cite{hopking} of changing the point of view but goes one step further in terms of computational efficiency and variance reduction.
Instead of equation \eqref{eq1}, one can  consider this new equation
\begin{equation}
\underset{  K\times 1}{P(S)} =  \underset{  K\times M}{P(S|D)} \underset{M\times 1}{P(D)}
\label{eq2}
\end{equation}
where now $P(S|D)$ is a $  K\times M$ matrix of conditional probabilities whose elements $P(S=S_k|D=D_i)$ represent the frequency of a particular stem $S_k$ given the set of texts which actually express the opinion $D=D_i$. 
Then, the solution of the problem is as follows
\begin{equation}
\text{(inverse problem)}\hspace{0.5cm}
\underset{M\times 1}{P(D)} = \underset{M\times M}{[P(S|D)^T P(S|D)]}^{-1} \underset{M\times   K}{P(S|D)}^T \underset{  K\times 1}{P(S)}
\label{eq3}
\end{equation}
Equation \eqref{eq3} is such that the direct estimation of the distribution of opinion $P(D)$ is obtained but individual classification is no longer possible. In fact, this is not a limitation as the accuracy of \eqref{eq3} with respect to \eqref{eq1} is vastly better (variance of estimates decreases from 15-20\% to 3-5\%). Moreover, researchers are comprehensibly more interested in the aggregate distribution of opinions throughout a population than in the estimation of individual opinion. For details see \cite{ceron_etal_2015}.

\subsection{Lead-lag estimation}\label{sec:leadlag}
When working with two or more time series, the usual need is to study their correlation but also potential causation effects: who leads who? This technique is called lead-lag estimation.
The lead-lag effect is a concept of common practice that has some history in financial econometrics and that we bring to social media analysis. In time series, for instance, this notion can be linked to the concept of Granger causality, and we refer to \cite{Comte96} for a general approach. From a phenomenological perspective, the lead-lag effect is supported by empirical evidence reported in \citet{Chiao04,deJong97} and \citet{Kang06}, together with \citet{Robert11} and the references therein.
The usual Granger-like approach has several limitations: 1) the time series must be of the same frequency (daily data, weekly date, etc);  2) testing for  causality often leads to discover a bidirectional effect; 3) linear time series are used (VAR or similar) to model the data.
An additional problem is that, if the frequency of time series increases, i.e. the lag between the data diminishes, the empirical correlation vanishes artificially due to the so-called {\it Epps} effect \citep{Zhang11}.

In our applications of Section \ref{sec:4}, the data have usually different frequencies, contain missing data and hence are also asynchronous, and there is no reason to assume a linear behaviour (in many cases seasonalities do exist).
To take into account all the mentioned features of the data, recently \cite{Hoffmann13} proposed a lead-lag estimator based on the Hayashi-Yoshida asynchronous covariance estimator \citep{hay-yos04,hay-yos05}. This estimator also overcome the Epps effect.

Let $\theta\in(-\delta, \delta)$ be the time lag between the two nonlinear processes $X$ and $Y$. Roughly speaking, the idea is to construct a contrast function $U_n(\theta)$ which evaluates the Hayashi-Yoshida estimator for the times series $X_{t}$ and $Y_{t+\theta}$ and then to maximize it as a function of $\theta$. The lead-lag estimator $\hat\theta_n$ of $\theta$ is defined as
$$
\hat\theta_n=\arg\max\limits_{-\delta < \theta<+\delta} |U_n(\theta)|.
$$
When the value of  $\hat\theta_n$ is positive it means that $X_t$ and $Y_{t+\hat\theta_n}$ (or $X_{t-\hat\theta_n}$ and $Y_{t}$) are strongly correlated, so we say ``$X$ leads $Y$ by an amount of time $\hat\theta_n$'', so $X$ is the {\it leader} and $Y$ is the {\it lagger}. Viceversa for negative  $\hat\theta_n$.

\section{Applications: linking SWBI to other well-being measures}
\label{sec:4}
In this section we present a few experiments that try to explain the relationship  between the SWBI  and  other well-being indexes and different types of health-related data. In particular, we will analyse the relationship between PM10 daily data in the province of Milano and the health components of SWBI; then we relate SWBI to the the weekly data about the incidence of influenza in Italy. The third analysis consists of the study of the correlation among the components of the BES index and those included in SWBI for different Italian regions, with annual data.

\subsection{The SWBI index}
\label{sec:4.0}
The SWBI index is the simple arithmetic mean\footnote{We use simple mean here for sake of simplicity: any reasonable and justified weighted mean of the eight indicators can be theoretically proposed as a synthetic well-being measure.} of the eight indicators $\tt emo$, $\tt fun$, $\tt rel$, $\tt res$, $\tt sat$, $\tt tru$,  $\tt vit$ and $\tt wor$ introduced in Section \ref{sec:2}.
Table \ref{tab:all} reports the yearly values of SWBI and its eight components. Data are available from February 2012 till November 2015 at the time of this writing. The analysis is based on a total of 143 millions of tweets posted in Italian and from Italy. 

\begin{figure}[t!]
\centering
	\includegraphics[height=0.4\textheight]{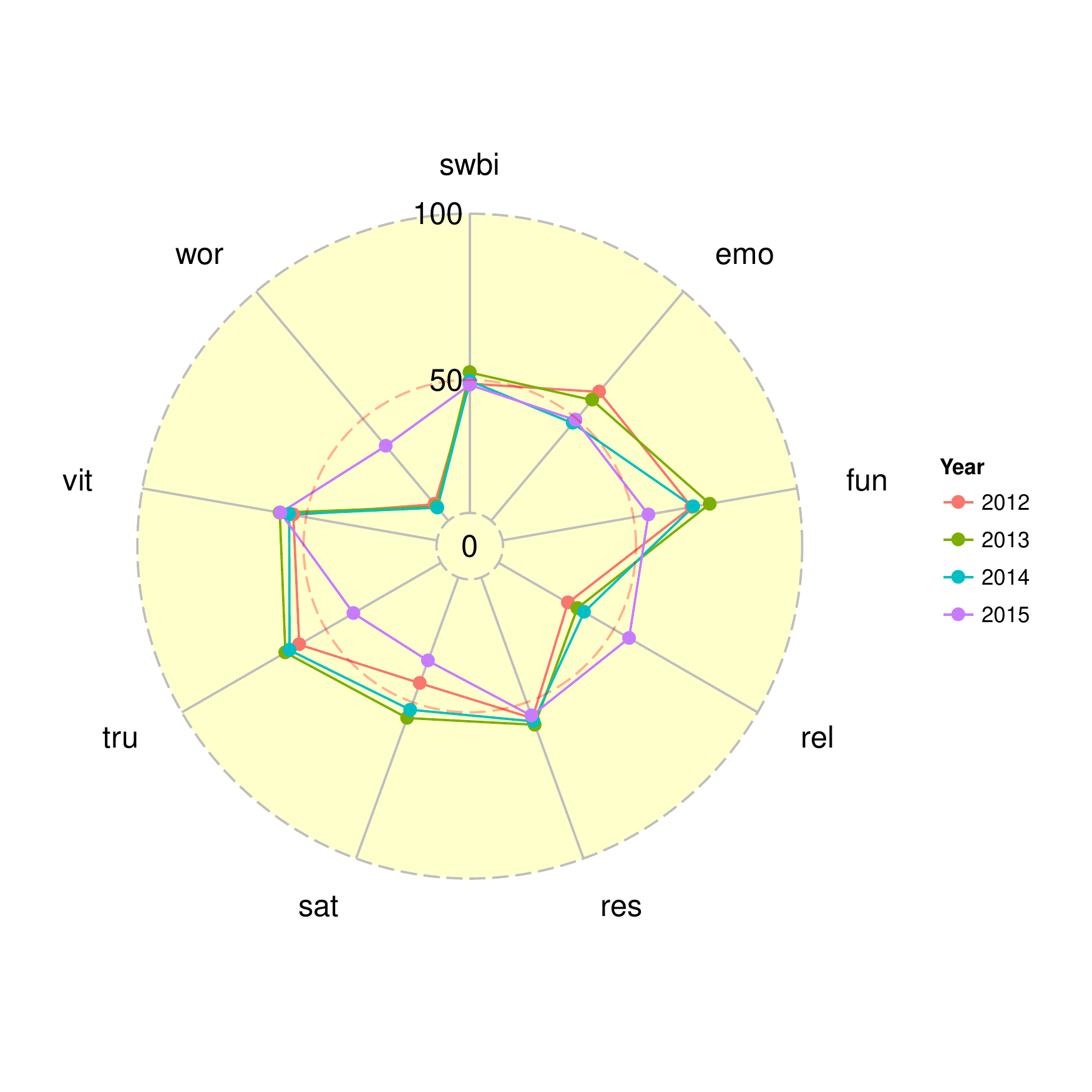}
\caption{\small Yearly average values of SWBI and its components. Data from Table \ref{tab:all}.}
\label{fig:radial}
\end{figure}

\begin{table}[ht]
\centering
\begin{tabular}{rrrrrrrrrr}
  \hline
 & SWBI & emo & fun & rel & res & sat & tru & vit & wor \\ 
  \hline
2012 & 48.87 & 60.55 & 67.76 & 34.10 & 55.10 & 43.88 & 59.22 & 53.91 & 16.44 \\ 
  2013 & 52.22 & 57.32 & 73.31 & 37.35 & 57.19 & 55.03 & 64.04 & 58.04 & 15.50 \\ 
  2014 & 49.69 & 48.24 & 68.26 & 39.73 & 56.11 & 52.37 & 62.59 & 55.15 & 15.10 \\ 
  2015  & 48.50 & 49.50 & 54.57 & 55.35 & 54.30 & 36.72 & 40.40 & 57.81 & 39.33 \\ 
   \hline
\end{tabular}
\caption{Average values of SWBI and its components.}
\label{tab:all}
\end{table}
It is interesting to notice that, if we look at the per capita GDP in Italy in 2012-2014 (data for 2015 are not available yet) and the value of the corresponding SWBI indicator we cannot find a clear common path, meaning that there is not necessarily a direct relationship between the level of economic activity of the country and the perceived well-being. The well-being indicator, in other terms, does not seem to simply reflect the conditions of the economic system, even in a period of serious economic crisis.  \begin{center}
\begin{tabular}{c|ccc}
Year & 2012 & 2013 & 2014\\
\hline
SWBI & 48.87& 52.22 & 49.69\\
\hline
GDP per capita (in euros, curr. prices) & 26760.0 & 26496.1 & 26545.8\\
\end{tabular}
\end{center}
\begin{figure}[t]
\centering
	\includegraphics[width=\textwidth,height=0.3\textheight]{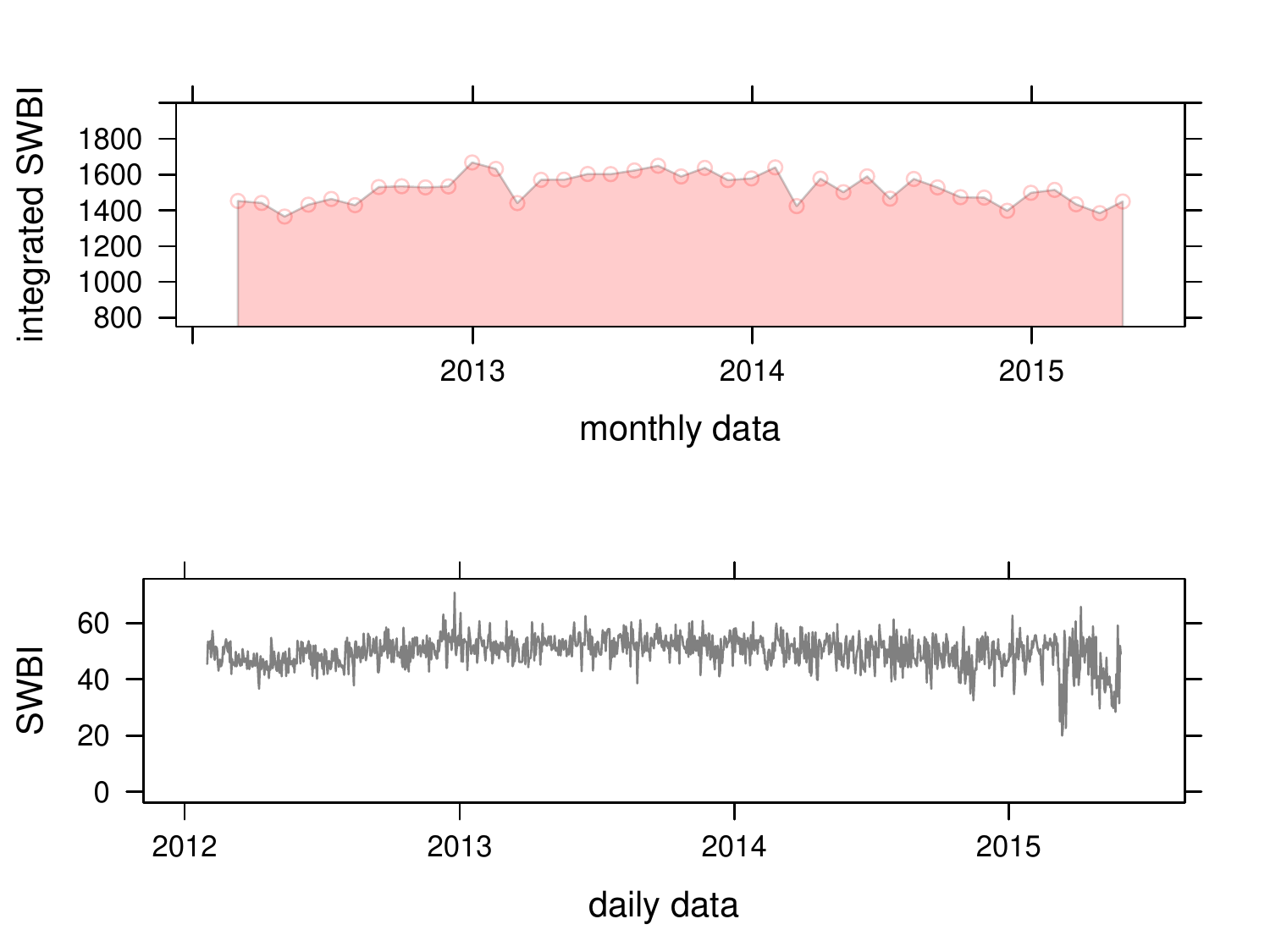}
\caption{\small daily values of SWBI (bottom panel) and its monthly integrated value (upper panel). The integrated value of SWBI represents the gross balance of well-being during each period.}
\label{fig:swbi}
\end{figure}
Figure \ref{fig:radial} represents the same data as Table \ref{tab:all}. It is easy to note that the values of the indicator in 2015 show remarkable differences (both positive and negative) compared to the trend of the previous years: see, in  particular, the increase in $\tt wor$ and $\tt rel$, or $\tt tru$, $\tt fun$ and $\tt sat$ for the opposite variation.
In addition to that, Figure \ref{fig:swbi} contains the plot of the daily values of SWBI (bottom panel) and its monthly integrated value (upper panel). The integrated value of SWBI represents the gross balance of well-being during each period. This representation dumps down the irregularity and high variability of daily estimates, which is typical in social media data. The above descriptive statistics need an in depth evaluation. It is only the case to note that the indicator registers both structural and volatile components of well-being and what we are showing is a preliminary and rough separation of the two, which is one of the discussion topic in the literature on well-being measurement. What follows represents a few examples of analyses related to health and other well-being measures.

\newpage
\subsection{The effect of PM10 pollutant abundance on well-being}
\label{sec:4.1}
\begin{figure}[t]
\centering
	\includegraphics[width=\textwidth,height=0.4\textheight]{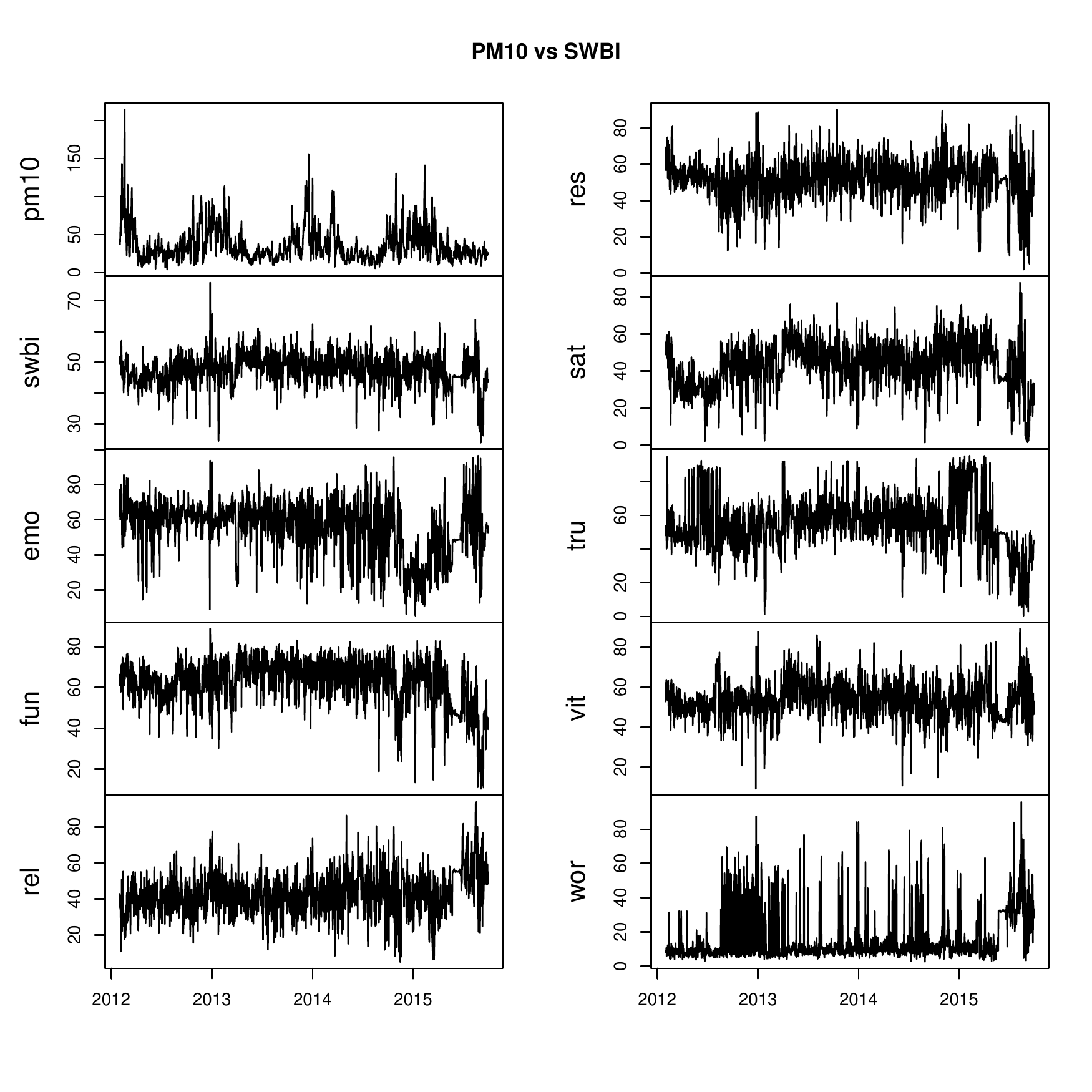}
\caption{\small Daily values of PM10, SWBI and its components for the city of Milan, from February 1st, 2012 to September 30th, 2015. Is there a leader among these time series?}
\label{zoo_MI}
\end{figure}

We consider data about the presence of PM10 in Milano. Data were downloaded from the regional agency for environmental protection (ARPA). We treated daily data from February 1st, 2012 to September 30th, 2015 which amounts to 2 millions tweets from Milan area in our data base, i.e. about 1.5 thousands per day. Figure \ref{zoo_MI} represents the time series of PM10, SWBI and its components. We run the  lead-lag analysis of Section \ref{sec:leadlag}  using a window of $\pm 5$ days to verify whether there is any leader among the dimensions of SWBI and the abundance of pollutant. It is not unlikely to expect that pollution may affect well-being,  possibly with some lag. 

\begin{table}[ht]
\begin{center}
{\small
\begin{tabular}{l|cccccccc}
  \hline
  &  $\tt emo$ & $\tt fun$ & $\tt rel$ & $\tt res$ & $\tt sat$ & $\tt tru$ & $\tt vit$ & $\tt wor$ \\ 
  \hline
 $\hat\theta_n$(PM10, $\cdot$) &    -4.1&  4.0&  5.0 &-4.1 & 1.0 & -1.1 & 5.0 & 3.0\\
corr(PM10, $\cdot$)   &-0.054 &-0.043 & 0.031 &-0.059& -0.068 &-0.037& -0.051 & 0.092 \\
   \hline
\end{tabular}
}
\end{center}
\caption{\small Lead-lag analysis: a positive value of $\hat\theta_n$ (value in days) means the time series of PM10 anticipates the time series in the column and viceversa. The second row of table reports the value of the Hayashi-Yoshida correlation estimator which is corrected for the Epps effect.}
\label{tab:llag}
\end{table}

The lead-lag analysis in Table \ref{tab:llag} shows the estimated lead-lag parameters in days.
From this analysis, it seems like the components {\it satisfying life} ($\tt sat$), {\it  vitality} ($\tt vit$), {\it working satisfaction} ($\tt wor$),  {\it positive functioning} ($\tt fun$), {\it relationship} ($\tt rel$) are anticipated by the amount of pollutant PM10 at different lags. In some cases, these indicators are negatively correlated with PM10.
This may indicate, for example, that increasing values of PM10 may cause a delayed effect on the perception of well-being in terms of overall satisfaction ($\tt sat$) and vitality ($\tt vit$). On the other hand, also the relationships between pollution and positive functioning ($\tt fun$) or job satisfaction ($\tt wor$) have an intuitive time direction, but they are likely a medium or long-run matter.
Again, this is far from estimating a causal effect, as pollution depends on many other aspects like temperature, humidity and traffic, but it seems that some evidence exists from these SWBI components, which deserves a more articulated and in-depth examination.

\subsection{Influenza and well-being}
\label{sec:4.2}
The Inter-universitary Research Centre on Influenza and other Transmissible Infections (CIRI-IT) is an Italian government institution whose task is research and scientific cooperation in the field of influenza and acute respiratory infections with particular regard to etiology, epidemiology, prevention and control. Among other activities, CIRI-IT publishes official data on the incidence of influenza in Italy. Weekly data from October 15, 2012 and May 23, 2015 are analyzed. This corresponds to 103 millions of tweets in our data base.
Similarly to the previous case of PM10 pollutant, we will make use of lead-lag estimation for daily data of SWBI against weekly data for influenza, therefore the two time series have different frequencies.
It turns out that the $\hat \theta_n({\tt flu}, {\tt SWBI}) = 4$ with a negative correlation of $corr({\tt flu}, {\tt SWBI}) = -0.062$ and the emotional component $\tt emo$ has a similar pattern, i.e. $\hat \theta_n({\tt flu}, {\tt emo}) = 1$ with $corr({\tt flu}, {\tt emo}) = -0.057$; similarly for $\tt wor$: $\hat \theta_n({\tt flu}, {\tt wor}) = 3$ with $corr({\tt flu}, {\tt wor}) = -0.066$. Intuitively, the number of flu cases increases the negative short-run feelings registered by the indicators, thus worsening the emotional component of well-being; at the same time, the sickness induced by flu affects job performance and reduces the perception of job satisfaction, also creating temporary difficulties in the work-life balance. On the whole, this yields a decrease in well-being (after 1 to 3 days), which shows its negative effect also on the aggregate SWBI indicator.

\subsection{BES versus SWBI}
\label{sec:4.3}
BES is the well-being index elaborated by the Italian Institute of Statistics (ISTAT) setting up from a dashboard of twelve dimensions: 
	 Environment ($\tt env$),
Health ($\tt hea$),
	Economic well-being ($\tt ewb$),
	 Education ($\tt edu$),
	 Work and lifetime ($\tt wlt$),	
	 Social relations ($\tt sre$),
	 Safety ($\tt saf$),
	 Subjective well-being ($\tt swb$),	
	 Landscape and culture ($\tt lac$),
	 Research and innovation ($\tt rin$),
	 Service quality ($\tt squ$),
	 Politics and institutions ($\tt poi$).

Although it is clear conceptually and statistically similar to the Better Life Index (BLI), BES does differ from it and other similar indexes presented in Section \ref{sec:1} in avoiding any form of aggregation: in the periodical report, for each dimension, the entire set of  proxies is presented and discussed. Even if ISTAT does not provide any form of aggregation inside and between dimensions composing BES, it is possible to find them on an on-line platform\footnote{\url{http://www.irespiemonte.it/iresinforma/index.php?option=com_content&view=article&id=37:qualita-della-vita-rapporto-bes&catid=5:societa&Itemid=3}} developed by IRES Piemonte (the Regional Institute for Economic and Social Research of Piemonte). IRES elaborates general and domain-specific composite indicators for all the Italian regions, presenting them with a format similar to BLI by OECD: avoiding any form of weighting but giving to the citizen the possibility of creating the index that reflects his/her own preferences. For our application we consider all the single dimensions of BES and SWBI, without summing them up in any weighted mean. Annual data from 2013 and 2014 are examined and the unit of analysis are the Italian regions. For the SWBI components we use data aggregated at regional level for a total of about 75 millions of tweets.

\begin{figure}[t]
\begin{center}
	\includegraphics[height=0.4\textheight]{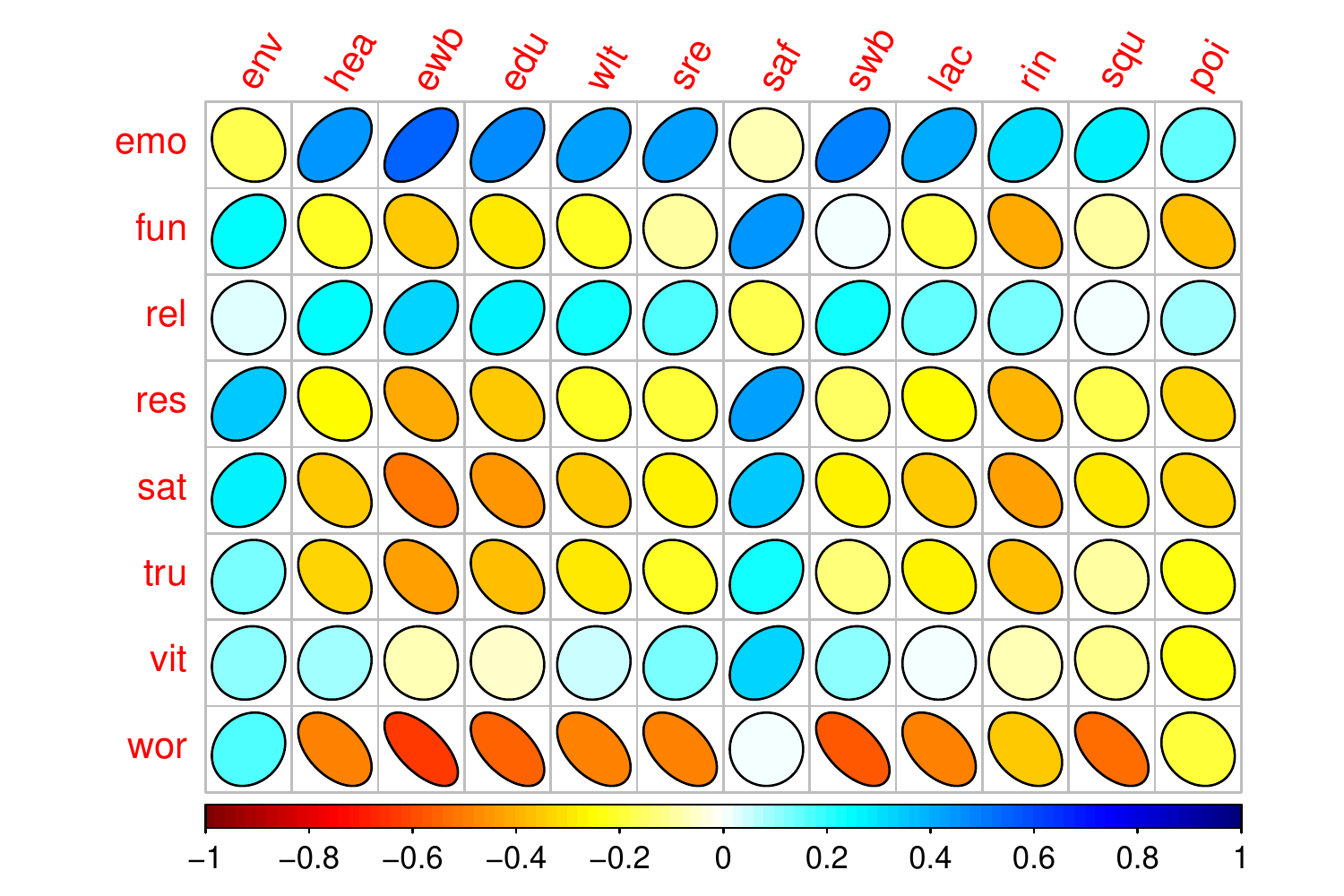}
\end{center}
\caption{\small Cross-Correlation matrix for: SWBI components vs BES variables. Increasing values are translated into colors from red (negative correlation) to blu (positive correlation).}
	\label{bes_cor_13}
\end{figure}

Our main aim is to understand whether a relationship exists between the BES indicators and the SWBI components. In other words: 
is the latent dimensions of well-being  common to the two sets of information?
The preliminary plot in Figure \ref{bes_cor_13} suggests that some correlations between the two set of variables exist. 

We apply a technique called \textit{canonical correlation analysis} (CCA) to extract, if any, the common latent dimension. 
Given two sets of variables $X$ and $Y$, CCA \citep{hotelling_1936} seeks for linear combinations of the variables of $Y$ that are maximally correlated to linear combinations of the variables of $X$. The analysis estimates the relationships and constructs new latent variables called canonical variables  \citep{legendre_2005}.
The canonical variates generated by CCA are orthogonal linear combinations of the original variables within each set $X$ and $Y$ that better explain the variability both within and between sets.
The Wilks's Lambda test suggests that only the first canonical dimension is significant. We use also the second in order to produce some dimension maps. Table \ref{cca} reports the canonical coefficients (loadings). The basic hypothesis we do is that there exists a latent dimension of well-being that is not observable and measurable, and that all dimensions that we observe are its proxy and therefore are related to it.
In general, when extracting a latent dimension, we expect a canonical dimension with the same relationship (positive or negative) with all the observed variables, in such a way that the latent dimension extracted is easy interpretable by a single polarity albeit by attributing different weights to the variables that compose it. If the polarity is positive, high values of the latent component indicate high well-being in the region and low values of the latent component indicate a low well-being in the region. 

\begin{table}[h!]
\centering
   \caption{Normalized Canonical Coefficients} 
   	\label{cca}
  \scriptsize{
\begin{tabular}{rrr}
  \hline
 & CanAxis1 & CanAxis2 \\ 
  \hline
  BES\\
  \hline
env & 0.43 & 0.14 \\ 
  hea & -0.32 & -0.20 \\ 
  ewb & -0.49 & -0.16 \\ 
  edu & -0.43 & -0.16 \\ 
  wlt & -0.49 & -0.19 \\ 
  sre & -0.28 & -0.06 \\ 
  saf & 0.36 & -0.12 \\ 
  swb & -0.29 & 0.11 \\ 
  lac & -0.46 & -0.26 \\ 
  rin & -0.69 & -0.28 \\ 
  squ & -0.73 & 0.01 \\ 
  poi & -0.58 & -0.16 \\ 
   \hline
   SWBI\\
     \hline
emo & -0.23 & 0.10 \\ 
  fun & 0.26 & 0.19 \\ 
  rel & -0.03 & 0.16 \\ 
  res & 0.30 & -0.05 \\ 
  sat & 0.39 & -0.04 \\ 
  tru & 0.15 & 0.29 \\ 
  vit & 0.24 & -0.23 \\ 
  wor & 0.47 & -0.23 \\ 
   \hline
\end{tabular}}
\end{table}

The canonical coefficients in the Table \ref{cca} have not all the same sign, even if it seems that the BES components are positively correlated with the latent dimension and the social well-being components are in general negative correlated. This means that the interpretation of the well-being latent dimension is not simple: there is no positive polarity of the indicators. This is clear also from Figure \ref{bes_cor_13}, where blue and red ellipses are shown. It is important to remember that the statistical units are the Italian regions and not the individuals; the social well being indicators have in fact been aggregated. The well-being in the different regions is characterized by several aspects. It can not be considered a unique one-dimensional that capture all aspects.

\begin{figure}[h!]
	\centering
	\includegraphics[width=\textwidth]{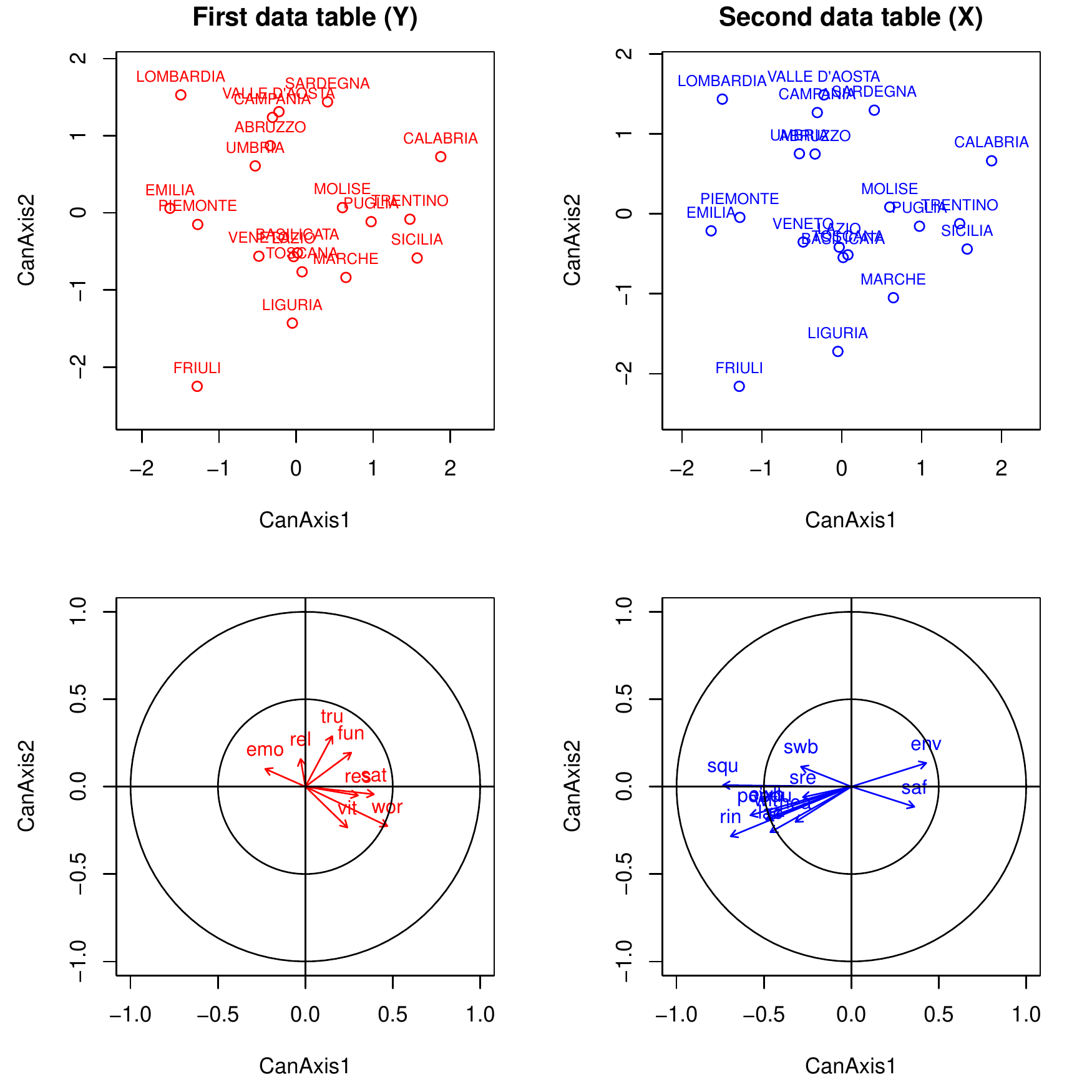}
	\caption{Biplots of the Canonical loadings and coefficients}
	\label{bbplot_cca13}
\end{figure}

Figure \ref{bbplot_cca13} shows the so-called biplot. The upper plots represent the scores for the first and the second canonical dimensions. In order to obtain the two sets of scores, the linear combination is calculated for each region using the loadings of the BES indicators and the loadings of the social well-being components. 
The representation of the Italian regions in the canonical axis (upper) is consistent either the indicators set is the BES (left) or the SWBI (right). This highlights that the latent variable actually depicts the same situation in both cases, i.e. that - conditional to the canonical coefficients - the two sets of indicators share the same piece of information.

The loadings maps (bottom) highlight that, at regional level, the BES dimensions can assume different directions and these directions correspond to different social dimension of well-being. This may imply that in some cases, despite the apparent similar content, the single BES or SWBI components measure different dimensions of well-being: this may be due, for example, to a difference between objective and perceived well-being evaluations.

Following this suggestion, we try to explain each SWBI dimension with the BES indicators. We run linear regression models of SWBI (as a simple mean of its components) and each of its components against the twelve dimensions of BES. Table \ref{reg2} reports the regression coefficients for each explanatory variable and their significance level.

\begin{landscape}
\begin{table}
{\tiny
\begin{tabular}{lcD{.}{.}{7}cD{.}{.}{7}cD{.}{.}{7}cD{.}{.}{7}cD{.}{.}{7}cD{.}{.}{7}cD{.}{.}{7}cD{.}{.}{7}cD{.}{.}{7}}
\toprule
&&\multicolumn{1}{c}{SWBI} && \multicolumn{1}{c}{emo} && \multicolumn{1}{c}{fun} && \multicolumn{1}{c}{rel} && \multicolumn{1}{c}{res} && \multicolumn{1}{c}{sat} && \multicolumn{1}{c}{tru} && \multicolumn{1}{c}{vit} && \multicolumn{1}{c}{wor}\\
\midrule
(Intercept)    &  & 50.523^{***} && 60.970^{***} && 68.443^{***} && 44.374^{***} && 53.869^{***} && 47.444^{***} && 58.896^{***} && 55.284^{***} && 14.907^{***}\\
               &  &  (1.270)     &&  (3.314)     &&  (2.941)     &&  (1.835)     &&  (1.347)     &&  (3.184)     &&  (1.571)     &&  (1.890)     &&  (0.786)    \\
env            &  &  -0.107      && -1.452^{***} &&  -0.326      &&  -0.037      &&   0.294      &&   0.527      &&   0.104      &&  -0.100      &&   0.136     \\
               &  &  (0.145)     &&  (0.379)     &&  (0.336)     &&  (0.210)     &&  (0.154)     &&  (0.364)     &&  (0.180)     &&  (0.216)     &&  (0.090)    \\
hea            &  &   0.003      &&   0.159      &&   0.115      &&  -0.112      &&   0.007      &&  -0.108      &&   0.091      &&   0.103      &&  -0.231     \\
               &  &  (0.216)     &&  (0.563)     &&  (0.499)     &&  (0.312)     &&  (0.229)     &&  (0.541)     &&  (0.267)     &&  (0.321)     &&  (0.134)    \\
ewb            &  & -0.668^{*}   &&  -1.206      && -1.622^{*}   &&  0.901^{*}   &&  -0.472      &&  -1.318      && -0.790^{*}   && -0.859^{*}   &&   0.022     \\
               &  &  (0.261)     &&  (0.680)     &&  (0.604)     &&  (0.377)     &&  (0.276)     &&  (0.654)     &&  (0.323)     &&  (0.388)     &&  (0.161)    \\
edu            &  &   0.049      &&   1.044      &&   0.807      &&  -0.582      &&   0.002      &&  -0.339      &&   0.071      &&  -0.450      &&  -0.164     \\
               &  &  (0.327)     &&  (0.853)     &&  (0.757)     &&  (0.473)     &&  (0.347)     &&  (0.820)     &&  (0.405)     &&  (0.487)     &&  (0.202)    \\
wlt            &  &  0.705^{***} &&  1.140^{*}   &&  1.644^{***} && -0.949^{**}  &&   0.360      &&  1.492^{**}  &&  0.902^{***} &&  0.816^{**}  &&   0.233     \\
               &  &  (0.184)     &&  (0.481)     &&  (0.427)     &&  (0.266)     &&  (0.195)     &&  (0.462)     &&  (0.228)     &&  (0.274)     &&  (0.114)    \\
sre            &  &   0.458      &&  -1.807      &&   0.188      &&   0.022      &&   0.746      &&  2.274^{*}   &&   0.382      &&  1.138^{*}   &&  0.724^{**} \\
               &  &  (0.358)     &&  (0.934)     &&  (0.829)     &&  (0.517)     &&  (0.379)     &&  (0.897)     &&  (0.443)     &&  (0.533)     &&  (0.222)    \\
saf            &  &  -0.199      &&   0.116      &&  -0.171      &&  -0.107      &&  -0.182      &&  -0.632      &&  -0.158      &&  -0.407      &&  -0.053     \\
               &  &  (0.185)     &&  (0.483)     &&  (0.429)     &&  (0.268)     &&  (0.196)     &&  (0.464)     &&  (0.229)     &&  (0.276)     &&  (0.115)    \\
swb            &  &   0.204      &&  1.862^{***} &&   0.707      &&  -0.337      &&  -0.220      &&  -0.390      &&   0.134      &&   0.239      && -0.360^{**} \\
               &  &  (0.187)     &&  (0.487)     &&  (0.432)     &&  (0.270)     &&  (0.198)     &&  (0.468)     &&  (0.231)     &&  (0.278)     &&  (0.116)    \\
lac            &  &  -0.363      && -1.829^{*}   &&  -0.984      &&   0.418      &&   0.152      &&  -0.034      &&  -0.199      &&  -0.404      &&  -0.021     \\
               &  &  (0.278)     &&  (0.726)     &&  (0.644)     &&  (0.402)     &&  (0.295)     &&  (0.697)     &&  (0.344)     &&  (0.414)     &&  (0.172)    \\
rin            &  &  -0.231      &&   0.146      &&  -0.316      &&  -0.045      && -0.429^{*}   &&  -0.699      &&  -0.168      &&  -0.153      &&  -0.182     \\
               &  &  (0.158)     &&  (0.411)     &&  (0.365)     &&  (0.228)     &&  (0.167)     &&  (0.395)     &&  (0.195)     &&  (0.235)     &&  (0.098)    \\
squ            &  &   0.206      &&   0.307      &&   0.192      &&   0.135      &&   0.216      &&   0.366      &&   0.034      &&   0.312      &&   0.091     \\
               &  &  (0.114)     &&  (0.298)     &&  (0.264)     &&  (0.165)     &&  (0.121)     &&  (0.286)     &&  (0.141)     &&  (0.170)     &&  (0.071)    \\
\midrule
R-squared      &  &     0.504    &&     0.615    &&     0.455    &&     0.517    &&     0.629    &&     0.600    &&     0.499    &&     0.519    &&     0.672   \\
adj. R-squared &  &     0.309    &&     0.464    &&     0.241    &&     0.327    &&     0.483    &&     0.443    &&     0.303    &&     0.330    &&     0.543   \\
sigma          &  &     1.164    &&     3.036    &&     2.695    &&     1.681    &&     1.234    &&     2.917    &&     1.439    &&     1.732    &&     0.721   \\
F              &  &     2.583    &&     4.070    &&     2.125    &&     2.723    &&     4.313    &&     3.818    &&     2.540    &&     2.746    &&     5.205   \\
p              &  &     0.021    &&     0.001    &&     0.053    &&     0.016    &&     0.001    &&     0.002    &&     0.023    &&     0.015    &&     0.000   \\
Log-likelihood &  &   -55.690    &&   -94.048    &&   -89.276    &&   -70.407    &&   -58.030    &&   -92.451    &&   -64.191    &&   -71.594    &&   -36.514   \\
Deviance       &  &    37.920    &&   258.116    &&   203.321    &&    79.152    &&    42.627    &&   238.306    &&    58.007    &&    83.990    &&    14.537   \\
AIC            &  &   137.379    &&   214.096    &&   204.551    &&   166.815    &&   142.059    &&   210.902    &&   154.382    &&   169.188    &&    99.028   \\
BIC            &  &   159.335    &&   236.052    &&   226.507    &&   188.770    &&   164.015    &&   232.857    &&   176.338    &&   191.143    &&   120.984   \\
N              &  &    40        &&    40        &&    40        &&    40        &&    40        &&    40        &&    40        &&    40        &&    40       \\
\bottomrule
\end{tabular}
}
\caption{\small Components of BES regressed against SWBI and its components.}
\label{reg2}
\end{table}
\end{landscape}

It is worth noting, first of all, that the SWBI has a negative relationship with the Economic well-being ($\tt ewb$) component of BES: as we previously pointed out, the SWBI does not sistematically reflects the economic conditions of the country. Moreover, the BES component which is most frequently related to SWBI and its sub-indicators is Work and lifetime ($\tt wlt$). The relationship with SWBI components is mostly positive (with the exception of ($\tt rel$), that may suggest a trade-off between satisfaction from work and the quality of friendship and family relations); it is quite disturbing, in this case, that the BES ($\tt wlt$) variable has a non-significant relationship with the SWBI ($\tt wor$), that is supposed to measure the same well-being dimension.

It is encouraging that the Subjective well-being BES component ($\tt swb$) is positively correlated to the emotional dimension of SWBI ($\tt emo$); at the same time the BES ($\tt swb$) is negatively related to the SWBI ($\tt wor$), again suggesting a sort of trade-off between quality of working time and overall perceived quality of life. On the other hand, it is quite convincing the positive and significant coefficient that connects the BES Social relations indicator ($\tt rse$) and the SWBI job satisfaction variable ($\tt wor$): maybe an absorbing job can worsen the perception of life quality, but al least it improves social relations.

Not easy to interpret the negative coefficient linking BES Environment ($\tt env$) variable to SWBI emotional component ($\tt emo$), and disappointing - to our current aims - that the coefficients of the BES indicator of Healt ($\tt hea$), though mostly positive, are never significant.

\section{Conclusion}
\label{sec:concl}
The eight-dimensional Social Well Being Indicator (SWBI) has been correlated to public health data with different frequency, exploiting the advantages of the lead-lag estimation procedure. 
Many components of the indicator show the expected behavior when correlated to the highest-frequency time series: PM10 pollutant abundance anticipates (i.e. has an impact on) several aspects of perceived well- being, in particular life satisfaction, vitality, job satisfaction and positive functioning. The same happens when SWBIU is correlated to weekly data on influenza: the number of flu cases increases the negative short-run feelings registered by the indicators, thus worsening the emotional component of well-being and negatively affecting job satisfaction. The impact is correctly registered, with a 4-days lag, by the overall SWBI indicator.

More fuzzy is the relationship between the components of SWBI and of the annual indicator BES: in fact, convincing correlation values can be found between the emotional component in SWBI and the subjective well-being dimension of BES, and significant correlation is also shown by the work and lifetime aspects, as measured by BES, and large part of SWBI sub-indicator. On the other hand, less clear results come from the canonical correlation analysis: SWBI and BES seem to share a significant latent variable, i.e. they carry, to some extent, the same piece of information about well-being. Nevertheless, the canonical coefficients of the two sets of indicators often show opposite signs, thus suggesting that the components of SWBI and BES, despite their apparent similar content, measure different dimensions or definitions of well-being.

\newpage

\bibliographystyle{spbasic}      
\bibliography{bibtex_SUPSI}   

\end{document}